\documentclass[11pt,a4paper]{article}
\usepackage[top=0.75in, left=2.5cm, right=2cm, bottom=1in]{geometry}
\usepackage[utf8]{inputenc}
\usepackage{amsmath}
\usepackage{graphicx}
\usepackage{booktabs}
\usepackage{mathptmx}
\usepackage{ntheorem}
\usepackage{tabularx}
\usepackage{longtable}
\usepackage{caption}
\usepackage{subcaption}
\usepackage{natbib}
\usepackage{multicol}
\usepackage{multirow}
\usepackage{tikz}
\usetikzlibrary{positioning,arrows.meta}
\usepackage{float} 
\usepackage{hyperref}
\hypersetup{colorlinks = true, linkcolor = blue, anchorcolor = blue, citecolor = blue, filecolor = blue, urlcolor = black}
\usepackage{bm}
\usepackage{authblk}

\theoremstyle{plain}
\theoremseparator{:}
\theoremheaderfont{\itshape}
\theorembodyfont{\normalfont}
\newtheorem{theorem}{Theorem}
\newtheorem{assumption}{Assumption}
\newtheorem{definition}{Definition}

\title{Identification and Estimation of Causal Estimands with Missing Not at Random Data}

\author[]{Faria Rauf Ria}
\author[]{Tarikul Islam}
\author[]{Mahbub A.H.M. Latif}

\affil[]{Institute of Applied Statistics and Data Science (IASDS), University of Dhaka, Dhaka 1000, Bangladesh}
\date{}

\begin{document}
\pagenumbering{arabic}

\maketitle
\thispagestyle{empty}
\begin{center}
 \textbf{Abstract} 
\end{center}

\noindent

Missing not at random (MNAR) data pose significant challenges for causal inference, particularly when both confounders and outcome are partially observed. Without additional assumptions beyond those required for causal inference, causal estimands are generally not identifiable under MNAR mechanisms. This paper first develops identification results for the causal estimand, the average treatment effect, under several plausible MNAR mechanisms using completeness conditions, and proposes an estimation approach based on the Expectation-Maximization (EM) algorithm. We further extend this identification and estimation framework to mediation analysis, enabling the estimation of natural direct and indirect effects under MNAR mechanisms. Through extensive simulation studies, we compare the proposed method with two widely used approaches for handling missing data, complete-case analysis and multiple imputation. The results show that the proposed method yields substantially lower bias under the considered MNAR mechanisms. Finally, we apply the proposed approach to NHANES data to estimate the causal effect of education on depression, with health condition as the mediator.

\vspace{0.5cm}

\clearpage
\section{Introduction}

In scientific research, causal inference allows researchers to assess whether a treatment of interest has a causal impact on an outcome. One widely used framework for estimating causal effects is the potential outcomes approach, which defines the outcomes that would have occurred under different treatment conditions as potential outcomes or counterfactuals \citep{holland1986statistics}. A causal estimand is a well-defined function of these counterfactuals that represents the causal effect of interest. A commonly used causal estimand is the Average Treatment Effect (ATE), or total effect, which quantifies the average causal effect of a treatment on an outcome. While the total effect captures the overall causal effect, causal mediation analysis decomposes it into a natural indirect effect (NIE), which acts through the mediator, and a natural direct effect (NDE), which acts independently of the mediator \citep{baron1986moderator, vanderweele2015explanation}. However, a significant challenge in the causal inference and mediation analysis arises from missing data \citep{bang2005doubly}. The mechanism of missingness can be categorized into three groups: (a) Missing Completely at Random (MCAR), in which the missingness is unrelated to all variables considered; (b) Missing at Random (MAR), in which conditional on the observed data, the missingness does not depend on unobserved variables; and (c) Missing Not at Random (MNAR) or non-ignorable missingness, in which missingness may still depend on unobserved variables even after conditioning on observed data \citep{little2019statistical}.

Prior causal inference studies have primarily focused on handling MAR mechanisms in confounders or outcome, although MNAR-type missingness has also been addressed to a more limited extent. \citet{ding2014identifiability} studied the identifiability of subgroup causal effects in randomized experiments when pretreatment covariates used to define subgroups were subject to nonignorable missingness. For ATE estimation under MNAR-type confounders in observational studies, \citet{yang2019causal} assumed an outcome-independent missingness mechanism and developed both a parametric likelihood-based estimator and a nonparametric two-stage least squares estimator.  \citet{sun2021semiparametric} later proposed a semiparametric alternative under the same outcome-independent missingness assumption. The challenge of MNAR confounders was further addressed by \citet{sun2025identification} under a treatment-independent missingness assumption. \citet{chen2023causal} addressed situations of self-censoring, in which the outcome itself affects its probability of being missing, and developed methods to obtain valid causal estimates under this non-ignorable missingness. \citet{zheng2023causal} focused on non-ignorable missing outcomes and introduced an auxiliary shadow variable that serves as a proxy for the missingness mechanism. More recently, identification assumptions for estimating average causal effects with missing not at random exposure and confounders have been proposed, along with efficient influence functions and doubly robust targeted maximum likelihood estimators (TMLE) for estimation \citep{wen2026estimating}.

However, in many practical settings, missing values may occur in both the confounder and the outcome, with the missingness mechanism being MNAR. For instance, consider estimating the causal effect of education on depression, where family wealth acts as a confounder influencing both variables. In this scenario, missing values may arise for both wealth and depression \citep{ettman2020wealth}. It is plausible that families with higher wealth might choose not to disclose their financial status, while individuals experiencing depression may be reluctant to report their mental health condition. Consequently, the missingness depends on unobserved values themselves, characterizing an MNAR mechanism. 

Similar challenges arise in causal mediation analysis, where missing values may occur in covariates, mediators, or outcomes \citep{dashtihandling, zhang2013methods, qin2019multisite}. When the missingness mechanism is MNAR, identification of the natural indirect and direct effects requires additional assumptions \citep{ayoku2023mediation}. For example, when assessing the effect of participation in a job training program on the earnings, with vocational attainment serving as a mediator, there may be missing values in earnings or educational attainment. In such cases, individuals with lower earnings may be less likely to report their income, leading to a missing data mechanism that depends on the unobserved value itself \citep{zuo2024mediation}. The identification of direct and indirect effects in the presence of missing outcome has been considered by several studies \citep{huber2020direct, li2017identifiability}. The method proposed by \citet{li2017identifiability} required a variable that is associated with the outcome but conditionally independent of the outcome missingness. On the contrary, \citet{huber2020direct} relied on a variable that influences the missing pattern of the outcome while remaining unrelated to the outcome when conditioned on the observed covariates.

Subsequent studies by \citet{zuo2024mediation} and \citet{nguyen2024self} considered settings in which both the mediator and the outcome are MNAR, with the first study focusing exclusively on identification and the second addressing both identification and estimation. MNAR-type missingness may also be present in confounders, e.g., \citet{shan2024efficient} considered MNAR-type missingness in confounders and used a shadow variable to provide identification and non-parametric estimation. However, in practical studies, missingness in the confounder and outcome often occurs simultaneously. This further complicates identification because, besides modeling the missingness mechanism for each variable, the dependence between the two missingness processes must also be accounted for. Despite its practical relevance, the simultaneous presence of MNAR missingness in both the confounder and the outcome has not yet been thoroughly explored in the literature, leaving a notable gap in the existing literature.

The objectives of this study are to extend the identification and estimation framework for the ATE in settings where both confounders and outcome are subject to MNAR mechanisms. We further develop methods for the identification and estimation of the NIE and NDE in mediation analysis when both confounders and the outcome are simultaneously subject to not at random missingness. The performance of the proposed estimators is evaluated through simulation studies, and the methods are illustrated using a nationally representative survey dataset to estimate the total, direct, and indirect effects.

The article is organized as follows: Section 2 presents an overview of the causal estimands of interest. Section 3 describes the identification and estimation of the causal estimands under plausible missigness mechanism,
and Section 4 describes the simulation study to evaluate and compare their performance with other estimators. Section 5 applies the proposed EM based method to the National Health and Nutrition Examination Survey (NHANES) data, and Section 6 concludes the article with a discussion of the findings.

\section{Causal Estimands}

This section presents a brief description of the causal estimands of interest. Among the estimands considered in this study, the ATE is introduced first, followed by the NDE and NIE in the mediation setting.

\subsection{Average Treatment Effect}

Let $A$ denote a binary treatment, where $A = 1$ represents the treatment group and $A = 0$ represents the control group. Let $\bm{X}$ denote the vector of potential confounders, and let $Y$ denote the outcome, which may be continuous or binary. Each individual has two potential outcomes, $Y(1)$ and $Y(0)$, corresponding to the treatment levels $A = 1$ and $A = 0$, respectively. The primary estimand of interest is the  Average Treatment Effect (ATE), defined as
\begin{equation}\label{ate}
\tau = E\big[Y(1) - Y(0)\big].
\end{equation}

For continuous outcomes, $\tau$ represents the average difference in the mean outcome under treatment and control. For binary outcomes, since $E\{Y(a)\}=P\{Y(a)=1\}$, the ATE can be expressed on the risk difference scale as
\begin{equation*}
\tau
= P\big\{Y(1)=1\big\} - P\big\{Y(0)=1\big\}.
\end{equation*}

To estimate the ATE using data from observational studies, several key assumptions are required. First, the \textit{unconfoundedness} assumption, also known as \textit{conditional exchangeability}, posits that the potential outcomes are independent of treatment assignment given a vector of confounders: 
\begin{align*}
\big\{ Y(1), Y(0) \big\} \perp\!\!\!\perp A \,\vert\, \bm{X}.
\end{align*} 

Second, the \textit{positivity} condition, which asserts that for some $\epsilon > 0$, $\epsilon \leq P(A = 1 \,\vert\, \bm{X}) \leq 1 - \epsilon.$  Lastly, the assumption of \textit{consistency} implies that an individual’s observed outcome corresponds exactly to the potential outcome that would occur under their actual treatment i.e., $Y = Y(a)$ if $A = a$. Under these three assumptions, the ATE in Equation~\eqref{ate} can be expressed in terms of the observed data $(Y, A, \bm{X})$ as
\begin{align}\label{ate_obs}
\tau 
&= E\big[ Y(1) - Y(0) \big] \nonumber\\
&= E_{\bm{X}} \big[
E\{ Y \,\vert\, A=1, \bm{X} \}
-
E\{ Y \,\vert\, A=0, \bm{X} \}
\big],
\end{align}

where, for binary outcomes, $E(Y \,\vert\, A, \bm{X}) = P(Y=1 \,\vert\, A, \bm{X})$. The expression Equation~\eqref{ate_obs} is one of the approaches to estimate causal effects from an observational study, which is known as standardization or g-computation \citep{causal}.

\subsection{Natural Direct and Indirect Effects}

Mediation analysis aims to understand the mechanism through which an exposure influences an outcome, by decomposing the total causal effect into components attributable to a specific intermediate pathway. A variable that is intermediate in the causal process linking an exposure to an outcome is known as a mediator. The average causal effect of an exposure on an outcome is also known as the total effect, which can be decomposed into an indirect effect operating through the mediator and a direct effect operating independently of the mediator. Let $M$ denote the mediator, with all other notation remaining as previously defined. We assume that $\bm{X}$ includes all measured potential confounders of the exposure–mediator, exposure–outcome, and mediator–outcome associations.

Let $M(a)$ denote the potential mediator value for the treatment status set at $a$. The potential outcomes $Y(a)$ are also influenced by the mediator alongside the treatment in mediation analysis. Therefore, $Y(a,m)$ represents the potential outcome value that would be observed if the treatment were set to $a$ and the mediator to $m$. To define causal effects within the counterfactual framework, the assumption of consistency and composition is essential. The composition assumption asserts for treatment $A$ set to $a$, $Y(a)=Y(a,M(a))$. The consistency assumption states for $A=a$ and $M=m$, $Y=Y(a,m)$, and for $A=a$, $Y=Y(a)$ and $M=M(a).$ The causal estimand NIE denotes the average change in the potential outcomes when treatment status is fixed at $a=1$ while the mediator changes from $M(0)$ to $M(1)$. Conversely, the estimand NDE represents the average change in the potential outcomes when the mediator value is fixed at $M(0)$ while the treatment status changes from 0 to 1. These estimands are defined as
\begin{align*}
\text{NIE} &= E \big[ Y \big(1, M(1)\big) - Y \big(1, M(0)\big) \big] \\
\text{NDE} &= E \big[ Y \big(1, M(0)\big) - Y \big(0, M(0)\big) \big].
\end{align*}

For binary outcomes, the above definitions are interpreted on the risk difference scale, with each expectation replaced by the corresponding probability, $P\{Y(a,M(a^*))=1\}$. The ATE can be expressed as the sum of NDE and NIE as $\tau \text{ = NDE + NIE}$ \citep{pearl2001direct}. For the identification of NDE and NIE from observational data, the following assumption of sequential ignorability must be satisfied \citep{imai2010general}: 
\begin{align*}
\text{(i)}\quad & \big\{Y(a', m), M(a)\big\} \perp\!\!\!\perp A \,\vert\,  \bm{X} \\
\text{(ii)}\quad & Y(a', m) \perp\!\!\!\perp M(a) \,\vert\, A, \bm{X} ,
\end{align*}

where $P(M = m \,\vert\, A = a, \bm{X} = \bm{x}) >0$ and $P(A = a \,\vert\, \bm{X} = \bm{x})>0$ for $a \in \{0, 1\}$ and $m\in \mathcal{M}$ and $\bm{x}\in \mathcal{X}$.

The following identification formula is used when no missing data is present:
\begin{align}\label{iden_med}
E \big[ Y\big(a, M(a')\big) \,\vert\, \bm{X}=\bm{x} \big] 
&= \sum_{m} \big\{ E (Y \,\vert\, A = a, M = m,  \bm{X} = \bm{x}) 
\times P(M=m \,\vert\, A = a', \bm{X} = \bm{x}) \big\},
\end{align}

for $a \in \{0, 1\}.$ Using the identification in Equation~\eqref{iden_med}, the causal estimands can be written as:
\begin{align*}
\text{NDE} &= 
E\big[ Y(a, M(a')) \,\vert\, \bm{X}=\bm{x} \big] - 
E\big[ Y(a', M(a')) \,\vert\, \bm{X}=\bm{x} \big] \\
&= \sum_m \Big\{ \big( 
E[Y \,\vert\, A = a, M = m, \bm{X}=\bm{x}] - 
E[Y \,\vert\, A = a', M = m, \bm{X}=\bm{x}] 
\big) \\
&\quad \times P(M = m \,\vert\, A = a', \bm{X}=\bm{x}) \Big\},\\[1em]
\text{NIE} &= 
E\big[Y(a,M(a)) \,\vert\, \bm{X}=\bm{x}\big] - 
E\big[Y(a,M(a')) \,\vert\, \bm{X}=\bm{x}\big] \\
&= \sum_m \Big\{
\big( P(M = m \,\vert\, A = a, \bm{X}=\bm{x}) - P(M = m \,\vert\, A = a', \bm{X}=\bm{x}) \big) \\
&\quad \times E[Y \,\vert\, A = a, M = m, \bm{X}=\bm{x}] \Big\}.
\end{align*}

For binary outcomes, since the conditional expectation $E(Y \,\vert\, A, M, \bm{X})$ equals the conditional probability $P(Y=1 \,\vert\, A, M, \bm{X})$, the identification formulas apply directly, with conditional expectations interpreted as conditional probabilities. The above expressions identify the NDE and NIE using observational data under the assumptions of sequential ignorability, positivity, consistency, and composition.

\section{Identification and Estimation with Missing Confounder and Outcome}

The presence of missing data in key variables, such as the exposure, mediator, confounders, or outcome, complicates the identification of the joint distribution. Since causal estimands are functionals of the full-data joint distribution, identifying this distribution is essential for identifying the causal estimands. Methods for handling missing data attempt to recover the full-data distribution from the observed data under specific missingness mechanisms. In the following subsections, we present three missingness mechanisms under which identification of the ATE is achieved, followed by three corresponding missingness mechanisms under which the NDE and NIE are identified. The following definition introduces the completeness condition used throughout the identification results.

\begin{definition}\label{def:completeness}
A function $f(X,Y)$ is said to be complete in $Y$ if $\int g(X)f(X,Y)\,d\nu(X)=0$ implies that $g(x) \xrightarrow{\text{a.s.}} 0$, for any measurable function $g(X)$ \citep{inference}.
\end{definition}

\subsection{Average Treatment Effect}\label{sec:ate_missing}

\begin{figure}[H]
\centering

\begin{subfigure}[t]{0.32\textwidth}
    \centering
    \includegraphics[width=\textwidth]{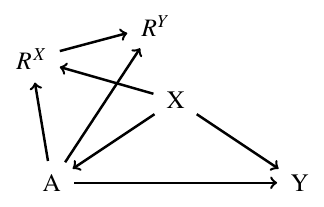}
    \caption{Assumption 1}
    \label{fig:mnar_assum1}
\end{subfigure}
\hfill
\begin{subfigure}[t]{0.32\textwidth}
    \centering
    \includegraphics[width=\textwidth]{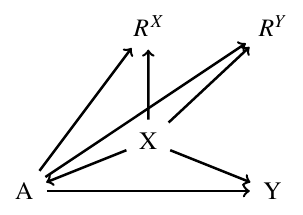}
    \caption{Assumption 2}
    \label{fig:mnar_assum2}
\end{subfigure}
\hfill
\begin{subfigure}[t]{0.32\textwidth}
    \centering
    \includegraphics[width=\textwidth]{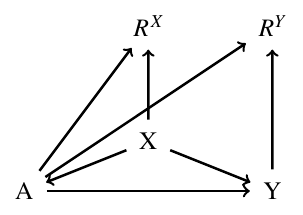}
    \caption{Assumption 3}
    \label{fig:mnar_assum3}
\end{subfigure}

\caption{
Causal DAGs representing the three MNAR mechanisms 
(a) Assumption 1: Conditional on $A$ and $X$, $R^X$ is independent of $Y$, and conditional on $(A,R^X)$, $R^Y$ is independent of $Y$.
(b) Assumption 2: Conditional on $A$ and $X$, $Y$, $R^X$, and $R^Y$ are mutually independent.
(c) Assumption 3: Conditional on $A$ and $Y$, $R^Y$ is independent of $X$ and $R^X$, and conditional on $A$ and $X$, $R^X$ is independent of $Y$ and $R^Y$.
}
\label{fig:mnar_dag_ate}

\end{figure}

Let $R^X$ and $R^Y$ denote the missingness indicators for the confounder $X$ and the outcome $Y$, respectively. In the case of MCAR, $(R^X, R^Y) \perp\!\!\!\perp (Y,A,X)$ whereas under MAR, $(R^X, R^Y) \perp\!\!\!\perp (Y,X) \,\vert\, A$. MNAR-type missingness can arise in various ways, for instance when $R^X$ depends on the unobserved value of $X$ or $Y$, since in both cases the probability of missingness depends on unobserved variables. Figure~\ref{fig:mnar_dag_ate} illustrates the three MNAR mechanisms considered in this paper under which identification is achieved.

Throughout the three mechanisms, we assume that $R^X \perp\!\!\!\perp Y \,\vert\, (A,X)$, allowing the missingness of the confounder to depend on its own value. These settings are particularly relevant for prospective studies, where treatment is measured before the outcome and missingness may occur in both the confounder and the outcome. In the following subsections, we establish the identification conditions and develop estimation procedures for the ATE under Mechanisms I, II, and III.

\subsubsection{Missingness Mechanism I}\label{sec:missing1}

The MNAR-type missingness mechanism I is described in the DAG (Figure \ref{fig:mnar_assum1}), where missing observations in the confounder can depend on itself, and missing observations in the outcome depend on missing observations in the confounder. Both types of missing observations depend on the treatment. In an observational study, researchers may investigate the effect of educational attainment ($A$) on individual earnings ($Y$), where family wealth ($X$) serves as an important confounder influencing both access to education and potential earnings. The missingness in family wealth ($R^X$) may depend on the actual wealth level and educational attainment. For example, individuals from wealthier households may be hesitant to disclose financial information, while those with lower levels of education may also be less likely to report their family wealth. Moreover, missing wealth data may occur alongside missing earnings data ($R^Y$), as individuals who refrain from disclosing wealth information might also avoid reporting their income. Under the DAG in Figure \ref{fig:mnar_assum1}, the following assumption can be made:

\begin{assumption}\label{assum1}
$(R^X, R^Y) \perp\!\!\!\perp Y \,\vert\, (A,X)$ and 
$R^Y \perp\!\!\!\perp (Y,X) \,\vert\, (R^X,A)$.    
\end{assumption}

Under Assumption~\ref{assum1}, together with the positivity and completeness conditions, the joint distribution $P(Y,A,X)$ is identifiable. The following theorem formalizes this result.

\begin{theorem}\label{theorem1}
Under Assumption~\ref{assum1}, $P(Y,A,X)$ is identifiable if $P(R^X=1, R^Y=1 \,\vert\, A,X) > 0$, $P(R^X=0, R^Y=1 \,\vert\, Y,A,X) > 0$ and $P(Y, A, X, R^X=1, R^Y=1)$ is complete in $Y$ for all $a$.
\end{theorem}

When $P(R^X=0, R^Y=1 \,\vert\, Y,A,X) =0$, the missingness of $X$ leads to the missingness of $Y$ and such missing data pattern is known as monotone missingness. In such a case, $P(Y, A, X)$ is not identifiable without additional assumptions. If we further assume $R^X \perp\!\!\!\perp X \,\vert\, A$ the missingness mechanism becomes MAR and $P(Y, A, X)$ is identifiable using complete cases. In this paper, however, we focus on the MNAR setting and therefore assume that $P(R^X = 0, R^Y = 1 \,\vert\, Y, A, X) > 0$. Under this missingness mechanism, identifying $P(Y, A, X)$ requires additional assumptions, which is described in Assumption~\ref{assum1}. We next provide details of the identification argument that ensures the recovery of the full-data distribution under the proposed MNAR mechanism.

\paragraph{Identification}\leavevmode\vspace{0.5em}\par

We elaborate below on the necessity of the completeness condition by considering the discrete case in which $X$ and $Y$ have $J$ and $K$ categories, respectively. The joint distribution can be expressed as 
$$P(Y, A, X) = P(A , X )\, P(Y \,\vert\, A , X ).$$
Under Assumption~\ref{assum1}, the identification of $P(Y  \,\vert\,  A , X)$ follows from
\begin{align}\label{eq:miss1_1}
P(Y  \,\vert\,  A , X)= P(Y \,\vert\, R^X = 1, R^Y = 1, A , X).     
\end{align}

To identify  $P(A , X )$, we can write the joint distribution as:
\begin{align}\label{eq:miss1_2}
P(Y, A, X)
&=\frac{P(Y, A, X, R^X = 1, R^Y = 1)}
{P(R^Y = 1 \,\vert\, A, R^X = 1) P(R^X = 1 \,\vert\, A, X)}.
\end{align}

The term $P(R^Y = 1 \,\vert\, A, R^X = 1)$ in Equation~\eqref{eq:miss1_2} is observed because it conditions only on quantities that are fully observed. The unobserved term in the Equation~\eqref{eq:miss1_2} is $P(R^X = 1 \,\vert\, A, X),$ which needs to be identified. Now we can write the observed data distribution, which involves the observed outcome $Y$ and excludes the unobserved confounder $X$, as follows:
\begin{align}\label{eq:miss1_3}
P(Y, A, R^X = 0, R^Y = 1)
= \frac{P(R^Y = 1 \;\vert\; A, R^X = 0)}{P(R^Y = 1 \,\vert\, A, R^X = 1)} 
\sum_x P(Y, A, X=x, R^X = 1, R^Y = 1) \zeta_{a}(x),
\end{align}

where 
\begin{align}\label{eq:zeta_ate}
\zeta_a(x)=
\frac{P(R^X = 0 \mid A = a, X = x)}
     {P(R^X = 1 \mid A = a, X = x)}.
\end{align}

The proof of Equation~\eqref{eq:miss1_3} is given in Supplementary Section S1. With the left-hand side being observed, the only unknown elements in the equation are ${\zeta_{a}(x)}$. Thus, we attain a system of linear equations with $\{\zeta_a(x) : x \in \mathcal{X}\}$ as the unknowns. In matrix notation, Equation~\eqref{eq:miss1_3} can be expressed as

\begin{equation*}
\begin{bmatrix}
P(Y = y_1, A=a, R^X=0, R^Y=1) \\
\vdots \\
P(Y= y_K, A=a, R^X=0, R^Y=1)
\end{bmatrix}
=
c_a
\Theta_{a} 
\begin{bmatrix}
\zeta_{a}(x_1) \\
\vdots \\
\zeta_{a}(x_J)
\end{bmatrix},
\end{equation*}

where 
$$
c_a=
\frac{P(R^Y=1 \mid A=a,R^X=0)}
     {P(R^Y=1 \mid A=a,R^X=1)},
$$

$$
\Theta_a=
\begin{bmatrix}
P(Y=y_1,A=a,X=x_1,R^X=1,R^Y=1) &
\cdots &
P(Y=y_1,A=a,X=x_J,R^X=1,R^Y=1)
\\
P(Y=y_2,A=a,X=x_1,R^X=1,R^Y=1) &
\cdots &
P(Y=y_2,A=a,X=x_J,R^X=1,R^Y=1)
\\
\vdots & \ddots & \vdots
\\
P(Y=y_K,A=a,X=x_1,R^X=1,R^Y=1) &
\cdots &
P(Y=y_K,A=a,X=x_J,R^X=1,R^Y=1)
\end{bmatrix}.
$$

To ensure the uniqueness of the solutions for $\{\zeta_a(x_j): j=1,\ldots,J\}$, the matrix $\Theta_a$ must be of full column rank, that is, $\operatorname{rank}(\Theta_a)=J$. This full-rank condition requires that $J \le K$, where $K$ denotes the number of categories of $Y$. This condition corresponds to the completeness condition in the discrete setting. The completeness condition is satisfied under several commonly used parametric models, including exponential family of distributions and a class of location-scale family of distributions \citep{newey2003instrumental, hu2018nonparametric}. Once $\zeta_a(x)$ is identified, $P(R^X=1 \mid A=a, X=x)$
is also identified. Consequently, the joint distribution $P(Y,A,X)$ can be expressed in terms of the observed data.

\paragraph{Estimation}\leavevmode\vspace{0.5em}\par

For the purpose of estimation, the complete data likelihood in this case using the DAG in Figure~\ref{fig:mnar_assum1} can be expressed as:
\begin{align*}
L_c(\pmb{\phi})=&\prod_{i=1}^{n} P(Y_i, A_i, X_i,R_i^X = r^X, R_i^Y=r^Y; \pmb{\phi} )\\
    =&\prod_{i=1}^{n} 
    \Big[ P(Y_i\,\vert\, A_i, X_i; \pmb{\theta}) \,
      P(R_i^X=r^X \,\vert\, A_i, X_i; \pmb{\lambda})\,
      P(R_i^Y=r^Y \,\vert\, A_i,  R_i^X; \pmb{\eta}) \,
      P(A_i \,\vert\, X_i; \pmb{\gamma}) \,
      P(X_i\,\vert\, \pmb{\alpha}) \Big].
\end{align*}

Since $X$ and $Y$ have missing values, we compute the observed data likelihood by summing over all possible values of missing $X$ and $Y$. The observed data likelihood :
\begin{align}\label{eq:lik_obs}
L_{o}(\pmb{\phi})=
&\prod_{\{i: R^X_i = 1, R^Y_i = 1 \}} P(Y_i, A_i, X_i,R_i^X = 1, R_i^Y=1; \pmb{\phi} ) \notag \\
& \times \prod_{\{i: R^X_i = 0, R^Y_i = 1 \}} \sum_x P(Y_i, A_i, X_i=x,R_i^X = 0, R_i^Y=1; \pmb{\phi} ) \notag \\
& \times \prod_{\{i: R^X_i = 1, R^Y_i = 0 \}} \sum_y P(Y_i=y, A_i, X_i,R_i^X = 1, R_i^Y=0; \pmb{\phi} ) \notag \\
& \times \prod_{\{i: R^X_i = 0, R^Y_i = 0 \}} \sum_x \sum_y P(Y_i=y, A_i, X_i=x,R_i^X = 0, R_i^Y=0; \pmb{\phi}).
\end{align}

Since direct maximization of the observed-data likelihood in the above equation is challenging, we employ the Expectation--Maximization (EM) algorithm to obtain the maximum likelihood estimators (MLEs) by treating the missing values of $X$, $Y$, or both as latent variables \citep{dempster1977maximum}. In the E-step, the conditional expectation of the complete data log likelihood is computed with respect to the missing variables. This computation depends on the missing data pattern. Specifically, three cases arise: (i) $X$ is missing while $Y$ is observed, (ii) $Y$ is missing while $X$ is observed, and (iii) both $X$ and $Y$ are missing. For example, when both $X$ and $Y$ are binary and missing, the conditional expectation is given by
\begin{equation*}
E\big\{I(Y_i=y_i, X_i=x_i) \,\vert\,  R_i^Y=0, R_i^X=0,A_i;\pmb{\phi}\big\}
=
\frac{P(Y_i=y_i, A_i,X_i=x_i, R_i^Y=0, R_i^X=0;\pmb{\phi})}
{\sum_{x=0}^{1}\sum_{y=0}^{1}{P(Y_i=y_i, A_i,X_i=x_i, R_i^Y=0, R_i^X=0;\pmb{\phi})}}.
\end{equation*}

Thus, for each subject with both $X_i$ and $Y_i$ missing, the E-step computes the probabilities corresponding to the four possible combinations of the missing values, namely, $(X_i, Y_i) \in \{(0,0), (0,1), (1,0), (1,1)\}.$ In the M-step, four copies of the dataset are created for each such subject, with each copy corresponding to one of these possible combinations. The likelihood in Equation~\eqref{eq:lik_obs} is then maximized using the corresponding probabilities as weights.

However, when either $X$ or $Y$ is continuous, the E-step of the standard EM algorithm is generally not available in closed form because it requires integration over the missing continuous variable. Therefore, we adopt a fractional imputation approach to approximate the required conditional expectations within the EM framework \citep{kim2011parametric}. For example, when the confounder $X$ is binary and the outcome $Y$ is continuous and both missing, we first enumerate the possible realizations of $x_i$, and subsequently generate the fractionally imputed outcomes $y_i^{(1)}, \ldots, y_i^{(S)}$ corresponding to each realization of $x_i$ from a proposal distribution $h(Y_i\,\vert\,A_i, X_i=x_i)$. Then, we compute the fractional weight for each imputed observation. The Monte Carlo approximation of the conditional expectation becomes:
\begin{equation*}
E\big\{L_{fi}(Y_i=y_i)\mid R_i^Y=0, R_i^X=0,A_i, X_i=x_i;\pmb{\phi}\big\}
\approx
\sum_{s=1}^S L_{fi}(Y_i=y_i;\pmb{\phi}) \hat{w}(y_i^{(s)}),
\end{equation*}

where the fractional weight is defined as
\begin{equation*}
\hat{w}(y_i^{(s)})
\propto
\frac{P(Y_i=y_i^{(s)}, A_i,X_i=x_i, R_i^Y=0, R_i^X=0)}
{h(Y_i=y_i^{(s)} \,\vert \,A_i, X_i=x_i)}.
\end{equation*}

The fractional weights are normalized such that $\sum_{s=1}^{S}\hat{w}(y_i^{(s)})=1$. Once the fractional weights are computed, the conditional expectation can be approximated as a weighted sum over the imputed values. Iterating this procedure within an EM framework allows for the estimation of the model parameters $\bm{\phi}$. The estimated parameters are subsequently utilized in the corresponding identification formulas to estimate the ATE.

\subsubsection{Missingness Mechanism II}\label{sec:missing2}

We next consider MNAR-type missingness mechanism II, illustrated in Figure~\ref{fig:mnar_assum2}, where the missingness of both the confounder $X$ and the outcome $Y$ depends on the treatment $A$ and the confounder $X$. In the example presented in Section~\ref{sec:missing1}, Assumption~\ref{assum2} implies that missingness in earnings may depend on the household's wealth status, which is a reasonable assumption in practice. This mechanism is formally specified in the assumption that follows:

\begin{assumption}\label{assum2}
 $Y, R^Y,$ and $R^X$ are mutually independent given $A$ and $X$.
\end{assumption}

We define a random variable $Y^* = (YR^Y, R^Y)$, which will be utilized later to facilitate the nonparametric identification. When $Y$ is observed ($R^Y = 1$), the variable $Y^*$ takes the value $Y^* = (y, 1)$. The probability associated with this event is $P\{Y^* = (y,1)\} = P(Y = y, R^Y = 1)$ for each $y \in \mathcal{Y}$. When $Y$ is missing ($R^Y = 0$), the actual value of $Y$ is unavailable. In this situation, no matter what value $y$ we might hypothetically assign to $Y$, all that is observed is that the outcome is missing. Consequently, the probability of the variable $Y^*$ taking the value $(y,0)$ does not depend on $y$ itself and is given by $P(Y^* = (y,0)) = P(R^Y = 0).$ Under Assumption~\ref{assum2}, the following theorem establishes the identification result.

\begin{theorem}\label{theorem2}
Under Assumption~\ref{assum2}, $P(Y, A, X)$ is identifiable if $P(Y^*, A, X, R^X=1)$ is complete in $Y^*$ for all $a$ and $P(R^X=1, R^Y=1 \,\vert\,A,X )>0$.   
\end{theorem}

To explain the identification result in Theorem~\ref{theorem2}, we present derivations below and provide intuition for the underlying identification strategy. The identification of $P(Y \,\vert\, A,X)$ under Assumption~\ref{assum2} follows from
\begin{align}\label{eq:miss2_1}
P(Y  \,\vert\, A , X)
= P(Y \,\vert\, R^X = 1, R^Y = 1, A, X).
\end{align}

To identify $P(A, X)$, we write the joint distribution
\begin{align*}
P(Y,A,X)
&= \frac{P(Y,A,X,R^X=1,R^Y=1)}
{P(R^X=1 \,\vert\, A,X)\,P(R^Y=1 \,\vert\, A,X)}.
\end{align*}

We need to identify both $P(R^X = 1 \,\vert\, A, X )$ and $P(R^Y = 1 \,\vert\, A, X)$. The joint distribution when $Y$ observed and $X$ missing can be written as:
\begin{align}\label{eq:miss2_2}
P(Y, A, R^X = 0, R^Y = 1)
= \sum_x P(Y, A, X=x, R^X = 1, R^Y = 1) \zeta_{a}(x),
\end{align}
where $\zeta_{a}(x)$ is defined in Equation~\eqref{eq:zeta_ate} for $y \in \mathcal{Y}$. The proof of the above equation is given in Supplementary Section S1. Next we write,
\begin{align*}
P( A, X, R^X = 1, R^Y = 0)
= P(R^Y=0\,\vert\, A, X)\, P(R^X=1 \,\vert\, A, X) \, P(A,X).
\end{align*}

So we need to identify both $P(R^Y=0\,\vert\, A, X)$ and $P(R^X=1 \,\vert\, A, X)$. Next we express the data where both $X$ and $Y$ are missing as:
\begin{align}\label{eq:miss2_3}
 P( A, R^X = 0, R^Y = 0)
= \sum_x P( A, X=x, R^X = 1, R^Y = 0) \zeta_a(x),
\end{align}

where $\zeta_a(x)$ is defined in Equation~\eqref{eq:zeta_ate}. The proof of the above equation is presented in Supplementary Section S1. When $R^Y=1$, achieving uniqueness requires that the observed $Y$ varies sufficiently across the $X$ values (Equation~\eqref{eq:miss2_2}). This corresponds to the usual completeness condition in $Y$. For $R^Y = 0$, meaning that $Y$ is unobserved in Equation~\eqref{eq:miss2_3}, the resulting expressions still involve the unknown quantities $\zeta_a(x)$. To identify $\zeta_a(x)$ uniquely, the vector of probabilities $P(A, X, R^X=1, R^Y=0)$ must provide enough independent information. The intuition is that the systematic pattern of missingness in $Y$, which depends on the values of $A$ and $X$, provides indirect information that can be used to identify the unknown $\zeta_a(x)$. In practical terms, to uniquely solve for the $J$ unknown parameters $\{\zeta_a(x_j), j=1 \, \ldots, J\}$, we need at least $J$ linearly independent equations. These are provided by the observed quantities $P(A = a, X = x_j, R^X = 1, R^Y = 0)$ for each distinct confounder value $x_j$. Essentially, the variation in missingness across different $X$-levels generates a system of equations, and a unique solution requires that the number of independent equations matches the number of unknowns. Formally, this requirement is expressed as completeness in $Y^* = (Y,R^Y)$, meaning that the missingness pattern provides linearly independent constraints that allow $\zeta_a(x)$ to be uniquely determined even when $Y$ is not observed. The estimation procedure follows the same structure as that described under Assumption~\ref{assum1}, with the distinction that here, the model for $R^Y$ incorporates $X$ in place of $R^X$.

\subsubsection{Missingness Mechanism III}\label{sec:missing3}

The MNAR-type missingness mechanism III is described in the DAG (Figure \ref{fig:mnar_assum3}), where the missingness of $Y$ ($R^Y$) may depend on the outcome $Y$ and treatment $A$, but not on the confounder $X$ or the missingness of $X$ ($R^X$). Similarly, the missingness of $X$ ($R^X$) may depend on the confounder $X$ and treatment $A$, but not on the outcome $Y$ or the missingness of $Y$ ($R^Y$). For example, when assessing the causal effect of education on depression, family wealth serves as a confounder affecting both variables. In this context, missing values may occur in both wealth and depression \citep{ettman2020wealth}. The missingness in income may depend on the income itself, as individuals with very high or very low earnings may be reluctant to report their financial information. Similarly, the missingness in mental health outcomes may depend on the individual’s mental health status, since those experiencing severe depression or anxiety may be less willing to disclose their mental health condition. Under the DAG shown in Figure \ref{fig:mnar_assum3}, we make the following assumption:

\begin{assumption}\label{assum3}
$R^Y \perp\!\!\!\perp (X, R^X) \,\vert\, (Y,A)$ and 
$R^X \perp\!\!\!\perp (Y,R^Y) \,\vert\, (A,X).$  
\end{assumption}

The following theorem establishes the identification result.

\begin{theorem}\label{theorem3}
Under Assumption~\ref{assum3}, $P(Y,A,X)$ is identifiable if $P(Y,  A, X, R^X = 1, R^Y = 1)$ is complete in $Y$ for all $a$, $P(Y,  A, X, R^X = 1, R^Y = 1)$ is complete in $X$ for all $a$, and $P(R^X = 1, R^Y = 1 \,\vert\, Y, A, X) >0$.    
\end{theorem}

The details of the identification results in Theorem~\ref{theorem3} are presented below. Under Assumption 3, $P(Y \,\vert\,A,X)$ is not directly identifiable from the complete cases, and the joint distribution can be written as:
\begin{align}\label{eq:miss3_1}
&P(Y, A,X)
= \frac {P(Y,  A, X, R^X = 1, R^Y = 1)}
{P(R^Y = 1 \,\vert\, Y,  A)\, P(R^X = 1 \,\vert\, A, X )}.
\end{align}

We need to identify both $P(R^Y = 1 \,\vert\, Y, A)$ and $P(R^X = 1 \,\vert\, A, X)$ in Equation~\eqref{eq:miss3_1}, since these quantities depend on unobserved values of the outcome $Y$ and the confounder $X$, respectively. To identify $P(R^X = 1 \,\vert\, X, A)$ from the above equation, the joint distribution of the observed $Y$ and $A$ when $X$ is missing can be written as:
\begin{align}\label{eq:miss3_2}
&P(Y, A, R^X = 0, R^Y = 1)
= \sum_x P(Y,  A, X=x, R^X = 1, R^Y = 1)\zeta_{a}(x),
\end{align}

where $\zeta_{a}(x)$ is defined in Equation~\eqref{eq:zeta_ate} for each $y \in \mathcal{Y}$. The proof of Equation~\eqref{eq:miss3_2} is given in Supplementary Section S1. The left-hand side is observable, comprising only the observed $Y$. Hence, the uniqueness of the solutions for the unknowns ${\zeta_{a}(x)}$ requires that the distribution $P(Y, A, X, R^X = 1, R^Y = 1)$ be complete in $Y$. Analogously, $P(R^Y = 1 \,\vert\, Y, A)$ can also be identified by replacing $R^X$ with $R^Y$, as shown below: 
\begin{align*}
P( A, X, R^X = 1, R^Y = 0)
&= \sum_y P(Y = y, A, X, R^X = 1, R^Y = 1)\eta_{a}(y),
\end{align*}

where 
$$\eta_{a}(y) = \frac {P(R^Y = 0 \,\vert\, Y = y, A=a)} {P(R^Y = 1 \,\vert\, Y = y, A=a )}.$$
To ensure unique identification, similarly here $P(Y,  A, X, R^X = 1, R^Y = 1)$ needs to be complete in $X$. To satisfy both completeness conditions, $X$ and $Y$ needs to have the same dimension or the same number of categories. The estimation procedure follows the same structure as in Assumption~\ref{assum1}, except that in this case, the model for $R^Y$ includes $Y$ in place of $R^X$.

\subsection{Natural Direct and Indirect Effects}

\begin{figure}[H]
\centering

\begin{subfigure}[t]{0.32\textwidth}
    \centering
    \includegraphics[width=\textwidth]{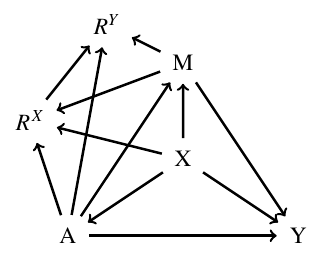}
    \caption{Assumption 4}
    \label{fig:mnar_assum4}
\end{subfigure}
\hfill
\begin{subfigure}[t]{0.32\textwidth}
    \centering
    \includegraphics[width=\textwidth]{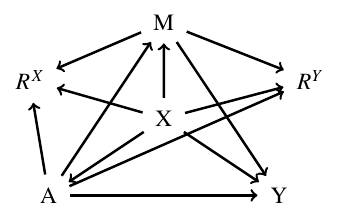}
    \caption{Assumption 5}
    \label{fig:mnar_assum5}
\end{subfigure}
\hfill
\begin{subfigure}[t]{0.32\textwidth}
    \centering
    \includegraphics[width=\textwidth]{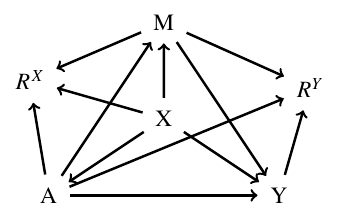}
    \caption{Assumption 6}
    \label{fig:mnar_assum6}
\end{subfigure}

\caption{
Causal DAGs representing the MNAR mechanisms
(a) Assumption~4: $R^X$ is conditionally independent of $Y$ given $(M, A, X)$, and
$R^Y$ is conditionally independent of $(X,Y)$ given $(R^X,A,M)$;
(b) Assumption~5: $R^X$ is conditionally independent of $(Y,R^Y)$ given $(M, A, X)$, and
$R^Y$ is conditionally independent of $(Y,R^X)$ given $(M,A,X)$;
(c) Assumption~6: $R^X$ is conditionally independent of $Y$ given $(M, A, X)$, and
$R^Y$ is conditionally independent of $(X,R^X)$ given $(Y,M,A)$.
}
\label{fig:mnar_dag_med}

\end{figure}

The assumptions considered in Section~\ref{sec:ate_missing} can be extended to accommodate missing not at random confounders and outcome for identifying the natural direct and indirect effects. Figure~\ref{fig:mnar_dag_med} illustrates the three MNAR-type missingness mechanisms (Mechanisms IV, V, and VI) considered in this setting.

Throughout Mechanisms IV--VI, the indicator variables $R^X$ and $R^Y$ retain their previous definitions as the missingness indicators for the confounder and the outcome, respectively. The assumptions and identification results corresponding to each mechanism are presented in the following subsections.

\subsubsection{Missingness Mechanism IV}\label{sec:missing4} 

Figure~\ref{fig:mnar_assum4} illustrates Mechanism IV, in which the missingness of $X$ may depend on $X$ itself, while the missingness of $Y$ may depend on whether $X$ is missing, but not directly on $Y$ or $X$, given the observed variables $A$ and $M$. Under the DAG shown in Figure \ref{fig:mnar_assum4}, the following assumption is made: 

\begin{assumption}\label{assum4}
$(R^X, R^Y) \perp\!\!\!\perp Y \,\vert\, (M,A,X)$ and 
$R^Y \perp\!\!\!\perp (Y,X) \,\vert\, (R^X,M,A)$.    
\end{assumption} 

To illustrate, consider the observational study in Section~\ref{sec:missing1}, where career satisfaction ($M$) mediates the effect of educational attainment ($A$) on individual earnings ($Y$). Family wealth ($X$) is an important confounder, and its missingness ($R^X$) may depend on the actual wealth level, educational attainment, and career satisfaction. Missing wealth data may also coincide with missing earnings data ($R^Y$), as individuals who do not disclose their family wealth may also avoid reporting their income. The identification result under Assumption~\ref{assum4} is stated in the following theorem.

\begin{theorem}\label{theorem4}
Under Assumption~\ref{assum4}, $P(Y,M,A,X)$ is identifiable if $P(R^X=1, R^Y=1 \,\vert\,Y,M, A,X) > 0$ for all $m,a,x$, $P(R^X=0, R^Y=1 \,\vert\, Y,M,A,X) > 0$ and $P(Y,M, A, X, R^X=1, R^Y=1)$ is complete in $Y$ for all $a,m$.    
\end{theorem}

To establish the identification result in Theorem~\ref{theorem4}, we follow the identification strategy developed for Mechanism I in Section~\ref{sec:missing1}. The identification details are provided in Supplementary Section S2. The complete and observed data likelihoods can be formulated similar to Assumption~\ref{assum1} using the DAG in Figure~\ref{fig:mnar_assum4}. Estimation is then carried out using the same approach described in Section~\ref{sec:ate_missing}.

\subsubsection{Missingness Mechanism V}

The MNAR-type missingness mechanism V is depicted in the DAG (Figure \ref{fig:mnar_assum5}), where the probability that $Y$ is missing may be influenced by the value of $X$. However, this missingness does not directly depend on the outcome $Y$ or the missingness status of $X$ ($R^X$) once we account for the variables $M$, $A$ and $X$. Under the DAG in Figure \ref{fig:mnar_assum5}, the subsequent assumption can be made:

\begin{assumption}\label{assum5}
$Y, R^Y,$ and $R^X$ are mutually independent given $M$, $A$, and $X$.    
\end{assumption}

Using the observational study introduced in Section~\ref{sec:missing4}, Assumption~\ref{assum5} assumes that missingness in earnings may depend on family wealth. This is a plausible assumption because individuals who withhold information about household wealth may also choose not to disclose their earnings. The identification under this assumption is established in Theorem~\ref{theorem5}.

\begin{theorem}\label{theorem5}
Under Assumption~\ref{assum5}, $P(Y,M, A, X)$ is identifiable if $P(Y^*, M, A, X, R^X=1)$ is complete in $Y^*$ for all $a,m$ and $P(R^X=1, R^Y=1, \,\vert\,M,A,X )>0$ .    
\end{theorem}

Following the identification strategy developed for Mechanism II in Section~\ref{sec:missing2}, the above theorem can be proved. The detailed derivation is provided in Supplementary Section S2. The estimation procedure is carried out using the approach described in Section~\ref{sec:ate_missing}.

\subsubsection{Missingness Mechanism VI} 

Figure~\ref{fig:mnar_assum6} illustrates MNAR-type missingness mechanism VI, where the probability that $Y$ is missing may be influenced by the value of $Y$ itself. However, this missingness does not directly depend on the confounder $X$ or the missingness status of $X$ ($R^X$) once we account for the observed variables $A$ and $M$. Under the DAG in Figure \ref{fig:mnar_assum6}, the subsequent assumption can be made:

\begin{assumption}\label{assum6}
$R^Y \perp\!\!\!\perp (R^X, X) \;\vert\; (Y,M,A)$ and 
$R^X \perp\!\!\!\perp (Y,R^Y) \;\vert\; (M,A,X).$     
\end{assumption}

For instance, in the previous example in mechanism IV, suppose the outcome is depression level. In that case, the missingness of depression is assumed not to depend on the missingness in wealth information. Rather, individuals with more severe depression may be less likely to report their mental health accurately. The following theorem provides the conditions required for identification of the joint distribution under the assumptions of Mechanism VI.

\begin{theorem}\label{theorem6}
Under Assumption~\ref{assum6}, $P(Y,M,A,X)$ is identifiable if $P(Y,M,  A, X, R^X = 1, R^Y = 1)$ is complete in $Y$ for all $a,m$, $P(Y, M, A, X, R^X = 1, R^Y = 1)$ is complete in $X$ for all $a,m$, and $P(R^X = 1, R^Y = 1 \,\vert\, Y,M, A, X) >0$.    
\end{theorem}

The identification result follows analogously to Mechanism III in Section~\ref{sec:missing3}, with details provided in Supplementary Section S2. The estimation procedure proceeds in a similar manner as discussed in Section~\ref{sec:ate_missing}.

\section{Simulation Study}

For the simulation study, we employed three methods to compare the performance in estimating the effects: (i) complete-case analysis (CC); (ii) multiple imputation by chained equations (MICE) with default settings (logistic regression for binary and predictive mean matching for continuous variables) \citep{van2011mice};  (iii) our proposed method based on EM algorithm. The performance of the estimators is evaluated using their bias and simulated SD. For an estimator of $\phi$ based on $S$ replications, the bias is computed as the average of $\hat{\phi}^{(s)} - \phi$ over $s = 1,\ldots,S$, and the simulated SD is given by the sample standard deviation of the values $\hat{\phi}^{(s)}$. We consider three scenarios defined by the type of confounder and outcome as shown in Table~\ref{tbl:scenario}.

\begin{table}[h!]
\centering
\caption{Scenarios defined by type of confounder and outcome}
\label{tbl:scenario}

\begin{tabular}{cll}
\toprule
Scenario & Confounder & Outcome \\
\midrule
$A$ & Binary & Binary \\
$B$ & Binary & Continuous \\
$C$ & Continuous & Continuous \\
\bottomrule
\end{tabular}
\end{table}

Across all assumptions and scenarios, the rate of missingness was specified to fall within the range of $20\%$ to $25\%$. We took sample sizes of $n \in \{500, 1000\}$, and replicated them $500$ times. Bias was obtained by averaging the bias values across the $500$ replications, and the simulated SD was calculated as the standard deviation of the estimates over those replications. Since the computational time was high, bootstrap was not used to estimate the standard error, although it can be applied.

\subsection{Simulation Study for ATE Estimation}

We consider two confounders, with one fully observed and the other subject to missingness. Let $X_{o}$ denote the fully observed confounder, generated from a standard Normal distribution, and let $X_{m}$ denote the confounder with missing values, generated as $X_{m} \sim \mathcal{N}(\mu, \sigma)$ if continuous, or $X_{m} \sim \text{Bernoulli}(\pi)$ if binary.
The binary exposure $A$ and the binary outcome $Y$ is generated from the following model:
\begin{align*}
\text{logit}\, P(A = 1 \,\vert\, X_{m}, X_{o}) &= \gamma_0 + \gamma_x X_{m} + \gamma_{c} X_{o}.\\
\text{logit}\, P(Y = 1 \,\vert\, A, X_{m}, X_{o}) &= \theta_0 + \theta_a A + \theta_x X_{m} + \theta_{c} X_{o}.
\end{align*}

When $Y$ is continuous, it is generated from:
\begin{align*}
Y \sim \mathcal{N}(\theta_0 + \theta_a A + \theta_x X_{m} + \theta_{c} X_{o}, 1).
\end{align*}

The indicator variable $R^X$ is generated according to the model:
\begin{align*}
\text{logit}\, P(R^X = 1 \,\vert\, A, X_{m}, C) &= \delta_0 + \delta_a A + \delta_x X_{m} + \delta_{c} X_{o}.
\end{align*}

The specification of the model for $R^Y$ differs across the Assumptions~\ref{assum1}, \ref{assum2}, and \ref{assum3}, with the corresponding models given below for each scenario.
\begin{align*}
\text{logit}\, P(R^Y = 1 \,\vert\, A, R^X, X_{o}) 
&= \lambda_0 + \lambda_a A + \lambda_{rx} R^X + \lambda_c X_{o} \nonumber \\
\text{logit}\, P(R^Y = 1 \,\vert\, A, X_{m}, X_{o}) 
& = \lambda_0 + \lambda_a A + \lambda_x X_{m} + \lambda_c X_{o} \nonumber \\
\text{logit}\, P(R^Y = 1 \,\vert\, A, Y, X_{o}) 
& = \lambda_0 + \lambda_a A + \lambda_y Y + \lambda_c X_{o}. \nonumber
\end{align*}

For generating the confounders, the parameter values are set as $\mu = 0.4$, $\sigma = 1$, and $\pi = 0.6$. The parameter values used for the other models are detailed in Supplementary Section S4. 

Table~\ref{tbl:sim_ate} reports the bias and SD for the CC, MI, and the proposed EM estimators under Assumptions~\ref{assum1}, \ref{assum2}, and \ref{assum3}. Overall for all scenarios and assumptions, the EM algorithm yields unbiased estimates compared to CC and MI. The only exception occurs under Assumption~\ref{assum3}, Scenario B, where $X$ is binary and $Y$ is continuous. In this setting, the differing dimensions of $X$ and $Y$ violate the completeness assumption, preventing the EM algorithm from fully recovering the joint distribution, which results in noticeable bias.

\begin{table}[h]
\centering
\caption{Bias and standard deviation (SD) for ATE estimation with MNAR confounder and outcome under Assumptions~1,2, and 3 across scenarios~$A$, $B$, and $C$.}
\label{tbl:sim_ate}

\begin{tabular}{c c c c c c c c c}
\toprule
 &  &  & \multicolumn{2}{c}{CC} & \multicolumn{2}{c}{MI} & \multicolumn{2}{c}{EM} \\
\cmidrule(lr){4-5} \cmidrule(lr){6-7} \cmidrule(lr){8-9}
Assumption & Scenario & $n$ & Bias & SD & Bias & SD & Bias & SD \\
\midrule

\multirow{6}{*}{1}
& $A$ & $500$  & $-0.051$ & $0.098$ & $-0.024$ & $0.098$ & $-0.009$ & $0.095$ \\
&   & $1000$ & $-0.048$ & $0.072$ & $-0.022$ & $0.072$ & $-0.006$ & $0.078$ \\
& $B$ & $500$  & $0.112$ & $0.101$ & $0.101$ & $0.101$ & $-0.008$ & $0.093$ \\
&   & $1000$ & $0.111$ & $0.084$ & $0.088$ & $0.084$ & $-0.001$ & $0.085$ \\
& $C$ & $500$  & $0.118$ & $0.093$ & $0.119$ & $0.093$ & $0.005$ & $0.089$ \\
&   & $1000$ & $0.122$ & $0.071$ & $0.125$ & $0.071$ & $0.004$ & $0.071$ \\
\midrule

\multirow{6}{*}{2}
& $A$ & $500$  & $0.032$ & $0.087$ & $0.012$ & $0.089$ & $0.008$ & $0.092$ \\
&   & $1000$ & $0.030$ & $0.066$ & $0.010$ & $0.071$ & $0.005$ & $0.089$ \\
& $B$ & $500$  & $-0.111$ & $0.112$ & $0.092$ & $0.114$ & $-0.010$ & $0.108$ \\
&   & $1000$ & $-0.110$ & $0.099$ & $0.089$ & $0.098$ & $-0.004$ & $0.089$ \\
& $C$ & $500$  & $0.312$ & $0.119$ & $0.122$ & $0.115$ & $-0.004$ & $0.127$ \\
&   & $1000$ & $0.301$ & $0.096$ & $0.108$ & $0.095$ & $-0.001$ & $0.101$ \\
\midrule

\multirow{6}{*}{3}
& $A$ & $500$  & $0.021$ & $0.089$ & $0.022$ & $0.087$ & $-0.007$ & $0.088$ \\
&   & $1000$ & $0.018$ & $0.067$ & $0.020$ & $0.069$ & $-0.003$ & $0.067$ \\
& $B$ & $500$  & $0.148$ & $0.137$ & $0.114$ & $0.132$ & $-0.119$ & $0.131$ \\
&   & $1000$ & $0.145$ & $0.116$ & $0.113$ & $0.111$ & $-0.112$ & $0.127$ \\
& $C$ & $500$  & $-0.231$ & $0.133$ & $-0.118$ & $0.101$ & $0.003$ & $0.110$ \\
&   & $1000$ & $-0.230$ & $0.112$ & $-0.120$ & $0.092$ & $0.002$ & $0.081$ \\
\bottomrule
\end{tabular}
\end{table}

\subsection{Simulation Study for Natural Direct and Indirect Effect Estimation}

For the simulation study with a mediator, the variables $A$, $X_{m}$ and $X_{o}$ are generated according to the models described previously. Let the variable $M$ is binary and simulated from the following model:
\begin{align*}
\text{logit}\, P(M = 1 \,\vert\, A, X_{m}, X_{o}) &= \beta_0 + \beta_a A + \beta_x X_{m} +
\beta_c X_{o}.
\end{align*}

Then $Y$ is generated from the model:
\begin{align*}
\text{logit}\, P(Y = 1 \,\vert\, A, X_{m}, X_{o}) &= \theta_0 + \theta_m M + \theta_a A + \theta_x X_{m} + \theta_{c} X_{o} + \theta_{ma} M A,
\end{align*}
when it is binary, and from the model:
\begin{align*}
Y \sim \mathcal{N}(\theta_0 + \theta_m M + \theta_a A + \theta_x X_{m} + \theta_{c} X_{o} + \theta_{ma} M A, 1),
\end{align*}

when it is continuous. Next, we introduce missingness in the confounder $X_{m}$. A binary missingness indicator $R^X$ is simulated according to the model:
\begin{align*}
\text{logit}\, P(R^X = 1 \,\vert\, M, A, X_{m}, X_{o}) = \delta_0 + \delta_a A + \delta_m M + \delta_x X_{m} + \delta_c X_{o}.
\end{align*}

The models for $R^Y$ are generated separately for each scenario, corresponding respectively to Assumptions~\ref{assum4}, \ref{assum5}, and \ref{assum6}
\begin{align*}
\text{logit}\, P(R^Y = 1 \,\vert\, M, A, R^X, X_o) &= \lambda_0 + \lambda_m M + \lambda_a A + \lambda_{rx} R^X + \lambda_c X_{o} \nonumber \\
 \text{logit}\, P(R^Y = 1 \,\vert\,M, A, X_m, X_o) &= \lambda_0 + \lambda_m M + \lambda_a A + \lambda_x X_m + \lambda_c X_{o} \nonumber  \\
 \text{logit}\, P(R^Y = 1 \,\vert\,M, A, Y, X_o) &= \lambda_0 + \lambda_m M + \lambda_a A + \lambda_y Y + \lambda_c X_{o}. \nonumber 
\end{align*}

The specific parameter values used in the data-generating models are summarized in Supplementary Section S4. 

The simulation results for the natural direct and indirect effect estimates under CC, MI, and EM methods are presented in Table~\ref{tbl:sim_assum4},  Table~\ref{tbl:sim_assum5}, and Table~\ref{tbl:sim_assum6}, corresponding to Assumptions~\ref{assum4}, \ref{assum5}, and \ref{assum6}, respectively. Across all scenarios, the EM algorithm produces approximately unbiased estimates, whereas the CC and MI methods exhibit noticeable bias. A notable exception occurs under Assumption~\ref{assum6} in Scenario B, where the EM estimate exhibits increased bias, which we attribute to a violation of the completeness condition. This pattern is also observed in the estimation of the ATE without a mediator.

%% assumption 4
\begin{table}[h]
\centering
\caption{Bias and SD for MNAR confounder and outcome missingness in mediation analysis under Assumption $4$ across scenarios $A$, $B$, and $C$.}
\label{tbl:sim_assum4}
\begin{tabular}{c c c c c c c c c c c c}
\toprule
 &  &  & \multicolumn{2}{c}{CC} & \multicolumn{2}{c}{MI} & \multicolumn{2}{c}{EM} \\
\cmidrule(lr){4-5} \cmidrule(lr){6-7} \cmidrule(lr){8-9}
Scenario & $n$ & Effect & Bias & SD & Bias & SD & Bias & SD \\
\midrule
\multirow{6}{*}{$A$}
& \multirow{3}{*}{$500$} & NDE & 0.032 & 0.085 & 0.012 & 0.089 & 0.003 & 0.064 \\
&                       & NIE & 0.027 & 0.053 & -0.118 & 0.130 & 0.004 & 0.055 \\
&                       & TE  & 0.059 & 0.079 & -0.106 & 0.133 & 0.007 & 0.061 \\
\cmidrule(lr){2-9}
& \multirow{3}{*}{$1000$} & NDE & 0.025 & 0.049 & 0.011 & 0.077 & 0.002 & 0.042 \\
&                        & NIE & 0.033 & 0.036 & -0.120 & 0.104 & 0.005 & 0.032 \\
&                        & TE  & 0.058 & 0.045 & -0.109 & 0.102 & 0.007 & 0.038 \\
\midrule
\multirow{6}{*}{$B$}
& \multirow{3}{*}{$500$} & NDE & -0.072 & 0.148 & -0.008 & 0.025 & -0.003 & 0.179 \\
&                       & NIE & 0.183 & 0.195 & -0.018 & 0.039 & 0.001 & 0.129 \\
&                       & TE  & 0.111 & 0.168 & -0.026 & 0.036 & -0.002 & 0.141 \\
\cmidrule(lr){2-9}
& \multirow{3}{*}{$1000$} & NDE & -0.058 & 0.125 & -0.013 & 0.016 & 0.002 & 0.137 \\
&                        & NIE & 0.175 & 0.163 & -0.016 & 0.028 & 0.001 & 0.103 \\
&                        & TE  & 0.117 & 0.138 & -0.028 & 0.024 & 0.003 & 0.120 \\
\midrule
\multirow{6}{*}{$C$}
& \multirow{3}{*}{$500$} & NDE & 0.077 & 0.135 & -0.125 & 0.133 & 0.002 & 0.105 \\
&                       & NIE & -0.162 & 0.164 & 0.041 & 0.099 & -0.004 & 0.140 \\
&                       & TE  & -0.085 & 0.162 & -0.085 & 0.102 & -0.002 & 0.134 \\
\cmidrule(lr){2-9}
& \multirow{3}{*}{$1000$} & NDE & 0.064 & 0.105 & -0.121 & 0.114 & 0.005 & 0.093 \\
&                        & NIE & -0.151 & 0.134 & 0.045 & 0.087 & -0.001 & 0.114 \\
&                        & TE  & -0.087 & 0.129 & -0.076 & 0.090 & 0.004 & 0.108 \\
\bottomrule
\end{tabular}
\end{table}

%% assumption 5
\begin{table}[ht]
\centering
\caption{Bias and SD for MNAR confounder and outcome missingness in mediation analysis under Assumption 5 across scenarios $A$, $B$, and $C$.}
\label{tbl:sim_assum5}
\begin{tabular}{c c c c c c c c c c c c}
\toprule
 &  &  & \multicolumn{2}{c}{CC} & \multicolumn{2}{c}{MI} & \multicolumn{2}{c}{EM} \\
\cmidrule(lr){4-5} \cmidrule(lr){6-7} \cmidrule(lr){8-9}
Scenario & $n$ & Effect & Bias & SD & Bias & SD & Bias & SD \\
\midrule
\multirow{6}{*}{$A$}
& \multirow{3}{*}{$500$} & NDE & 0.022 & 0.082 & 0.008 & 0.091 & 0.006 & 0.062 \\
&                       & NIE & 0.028 & 0.051 & -0.114 & 0.132 & 0.004 & 0.053 \\
&                       & TE  & 0.050 & 0.077 & -0.105 & 0.135 & 0.010 & 0.059 \\
\cmidrule(lr){2-9}
& \multirow{3}{*}{$1000$} & NDE & 0.031 & 0.051 & 0.011 & 0.075 & 0.007 & 0.040 \\
&                        & NIE & 0.024 & 0.038 & -0.112 & 0.106 & 0.003 & 0.030 \\
&                        & TE  & 0.055 & 0.043 & -0.101 & 0.104 & 0.010 & 0.036 \\
\midrule
\multirow{6}{*}{$B$}
& \multirow{3}{*}{$500$} & NDE & -0.062 & 0.132 & -0.051 & 0.135 & -0.006 & 0.107 \\
&                       & NIE & 0.123 & 0.166 & -0.021 & 0.101 & -0.002 & 0.142 \\
&                       & TE  & 0.061 & 0.165 & -0.072 & 0.104 & -0.008 & 0.136 \\
\cmidrule(lr){2-9}
& \multirow{3}{*}{$1000$} & NDE & 0.072 & 0.108 & -0.118 & 0.116 & 0.002 & 0.091 \\
&                        & NIE & -0.155 & 0.137 & 0.035 & 0.089 & -0.001 & 0.116 \\
&                        & TE  & -0.083 & 0.132 & -0.083 & 0.092 & 0.001 & 0.110 \\
\midrule
\multirow{6}{*}{$C$}
& \multirow{3}{*}{$500$} & NDE & -0.055 & 0.151 & -0.008 & 0.022 & 0.005 & 0.182 \\
&                       & NIE & 0.169 & 0.197 & -0.012 & 0.041 & 0.002 & 0.131 \\
&                       & TE  & 0.114 & 0.172 & -0.020 & 0.038 & 0.007 & 0.143 \\
\cmidrule(lr){2-9}
& \multirow{3}{*}{$1000$} & NDE & 0.122 & 0.128 & -0.041 & 0.014 & -0.003 & 0.139 \\
&                        & NIE & -0.059 & 0.166 & 0.118 & 0.026 & -0.001 & 0.105 \\
&                        & TE  & -0.063 & 0.140 & -0.077 & 0.022 & -0.004 & 0.122 \\
\bottomrule
\end{tabular}
\end{table}

%% assumption 6
\begin{table}[ht]
\centering
\caption{Bias and SD for MNAR confounder and outcome missingness in mediation analysis under Assumption 6 across scenarios $A$, $B$, and $C$.}
\label{tbl:sim_assum6}
\begin{tabular}{c c c c c c c c c c c c}
\toprule
 &  &  & \multicolumn{2}{c}{CC} & \multicolumn{2}{c}{MI} & \multicolumn{2}{c}{EM} \\
\cmidrule(lr){4-5} \cmidrule(lr){6-7} \cmidrule(lr){8-9}
Scenario & $n$ & Effect & Bias & SD & Bias & SD & Bias & SD \\
\midrule
\multirow{6}{*}{$A$}
& \multirow{3}{*}{$500$} & NDE & 0.031 & 0.081 & 0.011 & 0.084 & 0.007 & 0.059 \\
&                       & NIE & 0.024 & 0.049 & -0.112 & 0.125 & 0.003 & 0.050 \\
&                       & TE  & 0.055 & 0.074 & -0.101 & 0.127 & 0.010 & 0.056 \\
\cmidrule(lr){2-9}
& \multirow{3}{*}{$1000$} & NDE & 0.022 & 0.045 & 0.008 & 0.072 & 0.006 & 0.038 \\
&                        & NIE & 0.028 & 0.033 & -0.114 & 0.099 & 0.004 & 0.028 \\
&                        & TE  & 0.050 & 0.041 & -0.105 & 0.097 & 0.010 & 0.034 \\
\midrule
\multirow{6}{*}{$B$}
& \multirow{3}{*}{$500$} & NDE & -0.071 & 0.142 & -0.009 & 0.021 & -0.062 & 0.172 \\
&                       & NIE & 0.177 & 0.188 & -0.014 & 0.034 & 0.101 & 0.124 \\
&                       & TE  & 0.106 & 0.161 & -0.023 & 0.031 & 0.039 & 0.135 \\
\cmidrule(lr){2-9}
& \multirow{3}{*}{$1000$} & NDE & -0.055 & 0.120 & -0.008 & 0.012 & 0.093 & 0.130 \\
&                        & NIE & 0.169 & 0.157 & -0.012 & 0.023 & 0.054 & 0.098 \\
&                        & TE  & 0.114 & 0.131 & -0.020 & 0.019 & 0.147 & 0.115 \\
\midrule
\multirow{6}{*}{$C$}
& \multirow{3}{*}{$500$} & NDE & 0.072 & 0.128 & -0.118 & 0.126 & 0.003 & 0.100 \\
&                       & NIE & -0.155 & 0.158 & 0.035 & 0.094 & -0.001 & 0.134 \\
&                       & TE  & -0.083 & 0.156 & -0.083 & 0.096 & 0.002 & 0.128 \\
\cmidrule(lr){2-9}
& \multirow{3}{*}{$1000$} & NDE & 0.059 & 0.100 & -0.115 & 0.108 & 0.004 & 0.088 \\
&                        & NIE & -0.144 & 0.127 & 0.039 & 0.082 & -0.002 & 0.109 \\
&                        & TE  & -0.085 & 0.122 & -0.076 & 0.085 & 0.002 & 0.103 \\
\bottomrule
\end{tabular}
\end{table}

\section{Application to NHANES Data}

Using data from the 2015–2016 National Health and Nutrition Examination Survey (NHANES), this study estimates the causal effect of educational attainment on depression severity. The study sample included $5719$ individuals aged 20 years and older. The exposure variable was education, operationalized as a binary measure where $A = 1$ represented high school completion or higher ($n$ = 4,355) and $A = 0$ represented less than high school education ($n$ = 1,364). The main outcome variable ($Y$) of interest was depression severity, measured using the Patient Health Questionnaire-9 (PHQ-9) score  \citep{kroenke2001phq}.  The PHQ-9 is a nine-item, well-validated self-report instrument used to assess how frequently depressive symptoms occurred over the two weeks preceding the survey. Each item corresponds to a diagnostic criterion for major depressive disorder and is measured on a four-point Likert scale ranging from 0 (“not at all") to 3 (“nearly every day"). The total PHQ-9 score provides a continuous measure of depression severity ranging from 0 to 27, with higher scores indicating greater symptom burden. Family size, race, gender, and income-to-poverty ratio were included as confounders, while self-reported health condition was treated as the mediator. The dataset contained missing values in both the PHQ-9 scores and the income-to-poverty ratio, with approximately 22\% of participants having incomplete data for one or both variables.

\begin{figure}[h]
    \centering
    \includegraphics[width=0.7\textwidth]{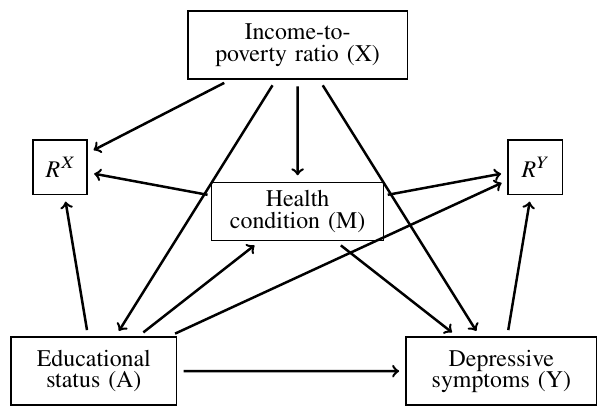}
    \caption{A DAG represents the relationship between education, health condition, income-to-poverty ratio, and depression. Family size, race, and gender were included as additional confounders but are omitted from the DAG for clarity.}
    \label{fig:mnar_data}
\end{figure}

Previous study showed that higher levels of education are associated with a lower prevalence of depressive symptoms \citep{li2022associations, bauldry2015variation}. However, such association-based analyses cannot establish causal effects. Therefore, we estimate both the ATE of education on depression severity and the natural direct and indirect effects to assess the mediating role of self-reported health condition. Figure~\ref{fig:mnar_data} presents the assumed causal framework used in the application. Education is the exposure, depression severity is the outcome, self-reported health condition is the mediator, and the income-to-poverty ratio is the partially observed confounder. Family size, race, and gender are also included as confounders in the analysis but are omitted from the figure for readability.

We estimate the ATE using the standardization approach in this scenario under Assumption~\ref{assum3}, as it represents the most plausible missing data mechanism in our study. The missingness in income-to-poverty ratio ($R^X$) is likely dependent on the variable itself, as higher-income families may be less willing to disclose financial information. Similarly, individuals experiencing depression may be less likely to report their mental health status. Crucially, conditional on income, the missingness in depression scores ($R^Y$) is independent of income values or their missingness patterns, and conversely, conditional on depression status, the missingness in income is independent of depression or its missingness. To draw inference under the standardization approach, bootstrap resampling is applied. After estimating the ATE, we investigate the mediating effect of perceived health condition on the effect of education on depression severity under Assumption~\ref{assum6}. 

\begin{figure}[H]
\centering
\includegraphics[width=1\textwidth, height=0.35\textheight]{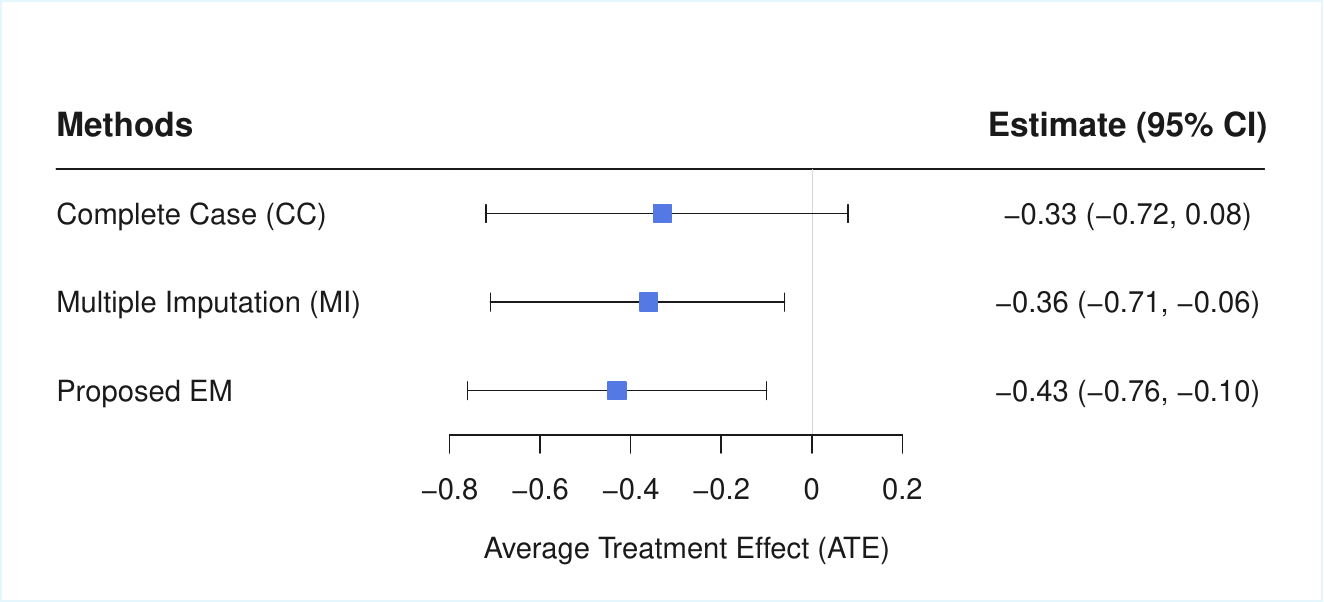}
\caption{Estimated ATE and 95\% CI for the effect of education on depression in the National Health and Nutrition Examination Survey.}
\label{fig:forest_ate}
\end{figure}

Figure \ref{fig:forest_ate} presents the estimated ATE and standard errors obtained from the three methods. Notably, there are meaningful differences in point estimates across methods, highlighting how missing data assumptions can substantially influence causal inference when both confounders and outcome are incomplete. Using the proposed EM algorithm, individuals with at least a high school education experience an average reduction of $0.43$ in the PHQ-9 score, which is larger than the effects observed with CC ($-0.33$) or MI ($-0.36$) analyses. Furthermore, the standard error from the EM-based estimator is smaller than those from CC and MI, suggesting that the proposed method provides more precise and potentially more accurate estimates by properly accounting for missingness in both confounders and outcome.

\begin{figure}[h]
\centering
\includegraphics[width=1\textwidth, height=0.35\textheight]{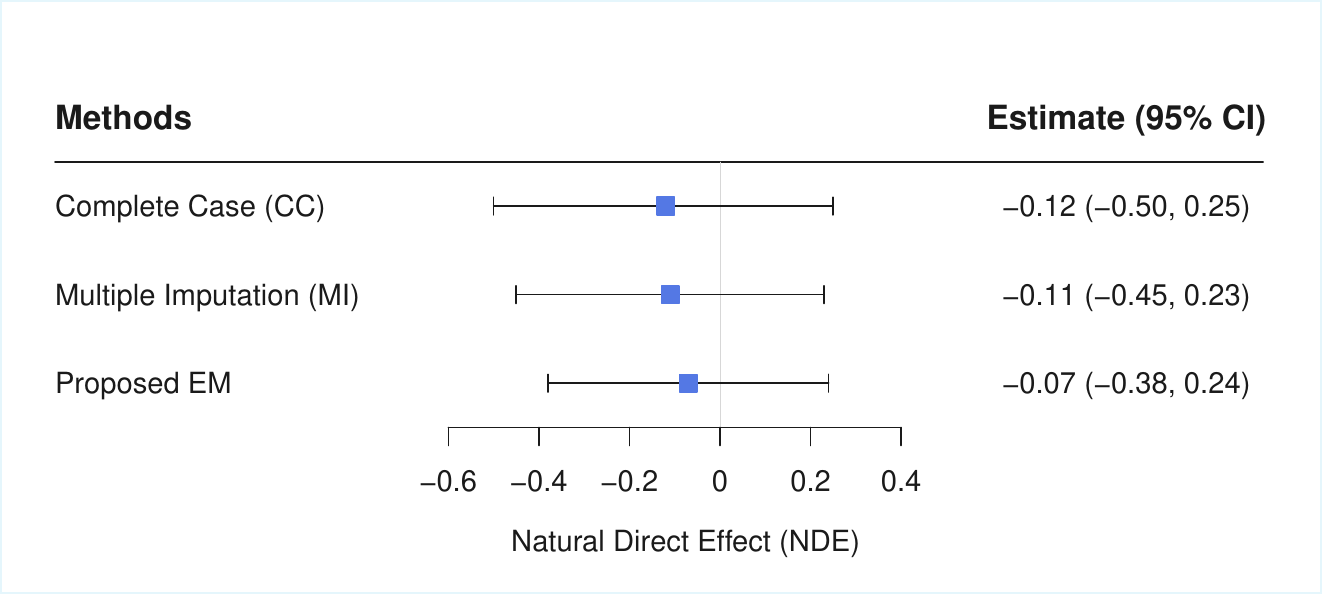}
\caption{Estimated Natural Direct Effect (NDE) and 95\% CI for the effect of education on depression in the NHANES, considering health condition as the mediator.}
\label{fig:forest_nde}
\end{figure}

\begin{figure}[h]
\centering
\includegraphics[width=1\textwidth, height=0.35\textheight]{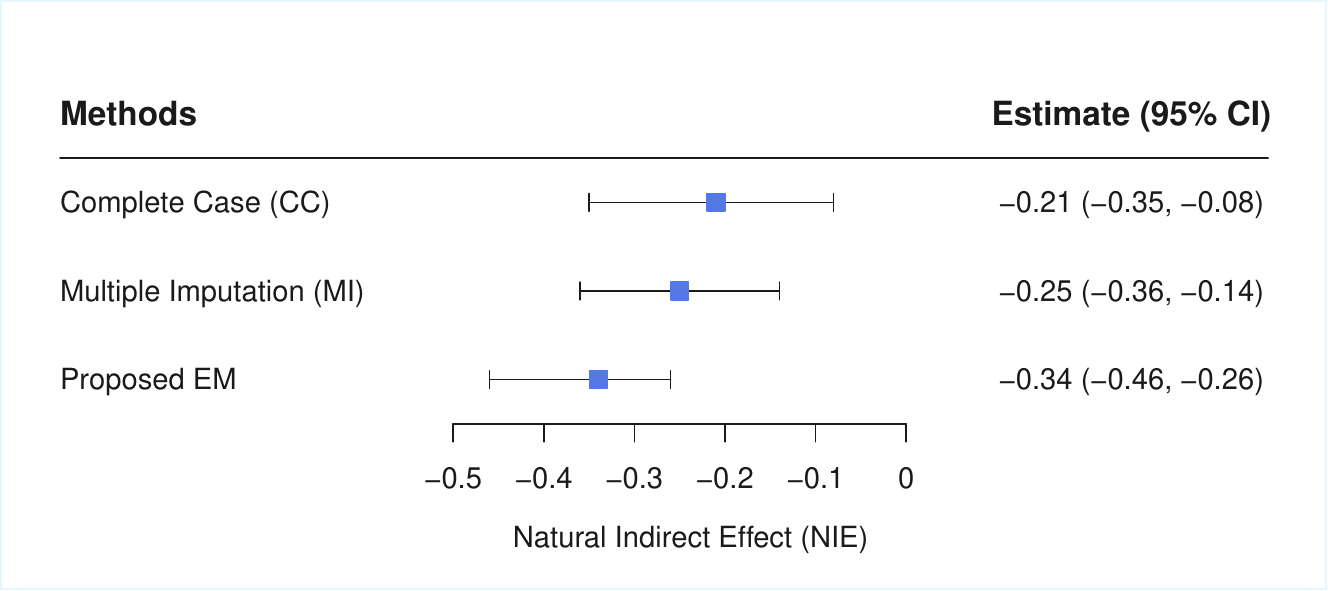}
\caption{Estimated Natural Indirect Effect (NIE) and 95\% CI for the effect of education on depression in the NHANES, considering health condition as the mediator.}
\label{fig:forest_nie}
\end{figure}

Figure \ref{fig:forest_nde} displays the estimated NDE and their 95\% CI, while Figure \ref{fig:forest_nie} displays the estimated NIE and 95\% CI obtained from the three methods. The NDE estimates indicate that, independent of health condition, higher education has a small negative effect on depressive symptoms, with values of $-0.12$ (CC), $-0.11$ (MI), and $-0.07$ (EM). In contrast, the NIE is substantially larger and negative across all methods, demonstrating that a considerable portion of education's protective effect on depression is mediated through improved health condition. Specifically, the NIE is estimated at $-0.21$ for CC, $-0.25$ for MI, and $-0.34$ for the EM algorithm. The CC estimator has the largest 95\% CI for both NDE and NIE. Importantly, the three methods produce different estimates. Despite differences in magnitude, all methods consistently indicate that education has a protective effect on depressive symptoms, primarily through health condition.

\section{Discussion}

A significant challenge in causal inference and mediation analysis is the pervasive issue of missing data. To obtain unbiased causal estimates, it is crucial to account for this missingness through appropriate statistical methods. A substantial methodological gap exists for Missing Not at Random (MNAR) data, where the missingness is dependent on the unobserved values of the variable. This work addresses the specific challenge of MNAR missingness in both confounders and the outcome by providing non-parametric identification strategy under plausible missingness mechanism. Building on the identification result, we construct parametric likelihood-based methods and derive the corresponding estimation procedure using the EM algorithm. 

We further extended the framework to handle MNAR-type missingness in mediation analysis, to identify and estimate the causal estimands, NIE and NDE. Although the analysis was illustrated using a single confounder for simplicity, the proposed framework is general and can accommodate missingness in multiple confounders. We implemented a simulation study to investigate the performance of the EM algorithm based proposed approach. The results revealed that while both CC and MI produced substantially biased estimates, the EM approach yielded unbiased results in both causal inference and mediation analysis. We applied our proposed causal estimation framework to NHANES data to illustrate its implementation.

Despite these contributions, the proposed framework has several limitations. First, the proposed framework is applicable only to the DAGs considered in this study and may not be directly applicable to more complex causal structures encountered in real-world settings. Second, the framework relies on completeness conditions for identification, which may not hold in all practical applications. In addition, the EM algorithm can be computationally demanding when applied to high-dimensional data.

One potential direction for future research is to extend the current fully parametric likelihood-based framework to a semiparametric estimating equation approach. Such estimating-equation based methods have been applied in settings where only the confounder exhibits non-ignorable missingness \citep{sun2025identification}; however, their use has not yet been extended to the more complex scenario in which both the confounder and the outcome simultaneously follow MNAR mechanisms. Investigating how this semi-parametric framework can be extended to accommodate simultaneous MNAR missingness in both variables constitutes an important direction for our future research.

\section*{Declaration of Conflicting Interests}
The authors declared no potential conflicts of interest with respect to the research, authorship, and/or publication of this article.

\section*{Funding}
The authors received no financial support for the research, authorship, and/or publication of this article.

\section*{Data Availability Statement}
The data that support the findings of this study are publicly available and can be downloaded from the Centers for Disease Control and Prevention (CDC) website at \url{https://www.cdc.gov/nchs/nhanes/}.

\bibliographystyle{apalike}
%\bibliography{ref}

\end{document}